\documentclass[reprint]{revtex4-1}
\usepackage{xspace} 
\usepackage{makecell} 
\usepackage{siunitx, mhchem}
\usepackage{color,soul}
\usepackage{graphicx}
\usepackage{bm}        
\usepackage{amssymb}   
\usepackage{amsmath,mathtools} 
\usepackage{breqn} 
\usepackage{multirow}
\usepackage{array}
\usepackage{booktabs}
\usepackage[export]{adjustbox}
\usepackage{textcomp}
\usepackage[normalem]{ulem} 
\newcommand{\mos}{MoS$_2$\xspace}
\newcommand{\ws}{WS$_2$ }

\begin{document}
	
	
\title{Tunability of electrical and thermoelectrical properties of monolayer \mos through oxygen passivation}
\author{Swarup Deb}
\email{swarupdeb@iitb.ac.in}
\author{Pritam Bhattacharyya}
\author{Poulab Chakrabarti}
\author{Himadri Chakraborti}
\author{Kantimay Das Gupta}
\author{Alok Shukla}
\author{Subhabrata Dhar}
\email{dhar@phy.iitb.ac.in}
\affiliation{Department of Physics$,$ Indian Institute of Technology Bombay$,$ Powai$,$ Mumbai 400076$,$ India}

\begin{abstract}
Electric and thermoelectric properties of strictly monolayer \mos films, which are grown using a novel micro-cavity based CVD growth technique, have been studied under diverse environmental and annealing conditions. Resistance of a thermoelectric device that is fabricated on a continuous monolayer \mos layer using photolithography technique has been found to reduce by about six orders of magnitudes upon annealing in vacuum at  525\,K. Seebeck coefficient of the layer also reduces by almost an order of magnitude upon annealing. When the sample is exposed to oxygen atmosphere, these parameters return to their previous values. In fact, it has been found that the electron concentration, mobility as well as the thermoelectric power of the material can be tuned by controlling the temperature of annealing and oxygen exposure. Once established, these values are maintained as long as the layer is not exposed to oxygen environment.  This can offer a unique way to control doping in the material provided an effective encapsulation method is devised. Such control is an important step forward for device application. The effect has been attributed to the passivation of di-sulfur vacancy donors present in the \mos film by physisorbed oxygen molecules. Band structural calculations using density functional theory have been carried out, results of which indeed validate this picture. 
\end{abstract}
\date{\today}
\maketitle

Two dimensional (2D) materials particularly transition metal dichalcogenides (TMDs), such as \mos, MoSe$_2$, \ws etc. have emerged as the materials for the next generation logic, electronic, opto-electronic devices and energy-related technologies \cite{aghosh,exciton_transistor,vacuum_desorption,energy_device}. Mechanical exfoliation\cite{dark_ex} and chemical vapour deposition (CVD)\cite{cvd_yale,h2s} are two most widely accepted techniques for preparing single layer 2D samples. Both of these techniques have certain positive and negative sides. For example,  exfoliated samples are superior in quality but they fail in terms of scalability. Typical size of monolayer regions in these flakes is not more than a few tens of micrometer\cite{Budania_2017,large_exfoliation}. Therefore, fabrication of large scale integrated circuits is not possible on such layers. On the other hand, CVD can provide monolayers with much larger area coverage\cite{cvd01,large_CVD,cvd00}. But, these films often suffer from high density of defects and grain boundaries\cite{polycrys_MoS2_GB,polycrys_MoS2_GB_review}. Irrespective of the preparation technique, increased surface to volume ratio has made these materials vulnerable to the ambiance. It should be noted that chalcogen vacancies, which act as donors, are omnipresent in these materials\cite{defectdop1,defectdop2}. In principle, these defects can influence the surface adsorption, which in turn can passivate these donors affecting both the concentration and the mobility of the carriers in the layer. If the density of the adsorbates, which passivate chalcogen vacancy donors, can be stabilized on the film surface by certain means, one can tune the doping level of the material, an important step towards device application. It should be mentioned that the optical properties of the film can also be affected by surface adsorption\cite{n2_defectPL}. In fact, there are recent studies on the effect of defect passivation on the optical properties in CVD grown monolayer TMD films\cite{2Dmat_PL,Redox_governed,prl,o2_PL_MIT}. However, the influence of adsorption on the electronic properties of monolayer TMDs has hardly been studied so far. 

Thermoelectrics have drawn a great deal of attention for the last several decades as they offer a nature-friendly way to convert heat to electricity and to use electricity for refrigeration. Low dimensional systems have come up as viable option\cite{gaas/algaas_TE,NW_TE,graphene_TE,InAs_TE}. Monolayer TMDs are predicted to be excellent thermoelectric materials because of their relatively high effective mass and the property of valley degeneracy\cite{thermoelectric1,prb_thermoelectric}. The Seebeck coefficient, which is one of the key parameter that decides the figure of merit of a thermoelectric material, has been experimentally found to be as high\cite{thermoelectric1} as 30\,mV\,K$^{-1}$ in 1L-\mos that is way more than any other nano-scaled materials\cite{graphene_TE,InAs_TE}. Yet, it should be mentioned that only a handful of experimental studies are carried out on thermoelectric properties of 1L-TMDs\cite{thermoelectric1,photothermoelectric,prb_thermoelectric}. It will be interesting to explore how the surface adsorption affects the Seebeck coefficient of these materials.

\begin{figure*}
	\centering
	\includegraphics[scale=0.3]{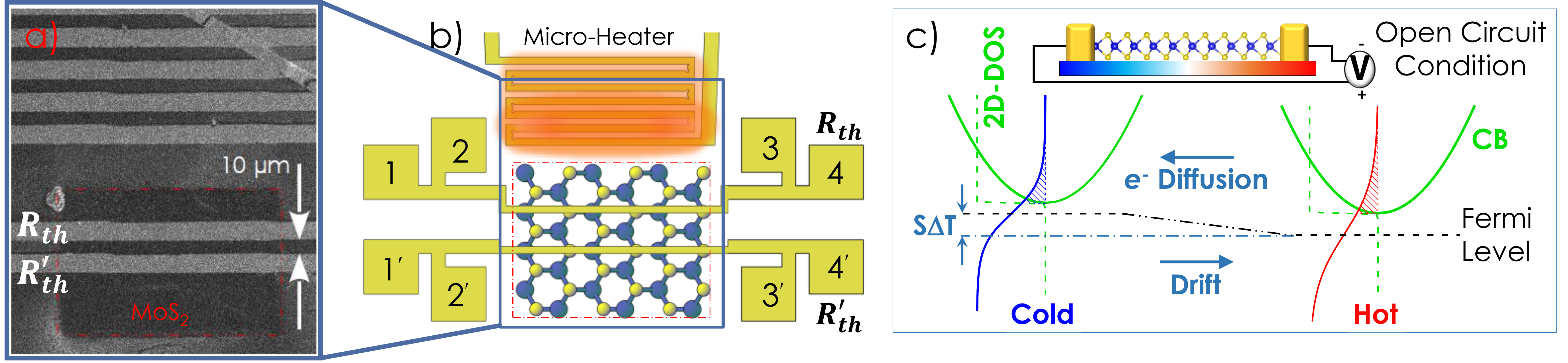}
	\caption{(a) Scanning electron micrograph of an actual device. (b) Schematic of the full device (not scaled). $R_{th}$ ($R_{th}^{\prime}$) is the four probe resistance of the metal line that connects contact pads 1 and 2 ($1^{\prime}$ and $2^{\prime}$) with 3 and 4 ($3^{\prime}$ and $4^{\prime}$). Heater dimension in the actual device is kept much larger than the length of the \mos region to ensure temperature uniformity across the cross-section. (c) Mechanism of charge flow and potential build up across the device in the presence of a thermal gradient for open circuit configurations is schematically illustrated. The (+) and (-) sign indicates the polarity of the voltmeter probes. In the given configuration measured thermoelectric voltage for n-type semiconductor will be $(-)ve$. }
	\label{Fig.1}
\end{figure*}

Recently, we have developed a novel micro-cavity based CVD growth technique, where strictly monolayer \mos (1L-\mos) film can be grown on c-sapphire substrates covering an area as large as a few mm$^2$\cite{pkm,sdeb}. These films show several G$\Omega$ of resistance in ambient conditions. Here, we have studied the thermoelectric properties of these films at different environmental conditions before and after annealing it in vacuum. A thermoelectric device is fabricated on a continuous film using the photolithography technique. It has been found that vacuum annealing at a temperature of about 525\,K can reduce the resistance of the device by about six orders of magnitudes and at the same time thermoelectric power also gets reduced by almost an order of magnitude. Interestingly, both the resistance ($R$) and the Seebeck coefficient ($S$) return to their previous values when the sample is exposed to oxygen environment. It has been observed that $R$, as well as $S$ of the device, can be tuned by controlling the temperature of annealing and oxygen exposure. These values hardly change as long as the device is not exposed to oxygen environment. The study attributes the effect to the passivation of sulfur vacancy donors by oxygen molecules, which are physisorbed on the surface. Band structural calculations within the framework of density functional theory have been performed to check the validity of the model. Theory shows that the adsorption of oxygen molecules at sulfur vacancy sites is indeed energetically stable. This results in the formation of energy levels 250~meV below the donor states arising from the sulfur vacancies. It has also been shown that these levels can capture electrons, which leads to the passivation of the donors.

Strictly monolayer \mos films were grown on \textit{c}-sapphire substrates using a microcavity based CVD technique. Prior to the growth, substrates were cleaned subsequently in TCE, acetone, and methanol and finally dipped in H$_2$O:HF (10:1) solution for 40\,sec. More details of the growth procedure have been discussed elsewhere\cite{pkm,sdeb}. Standard optical lithographic technique was used for device fabrication. Layers of two different metals \textit{viz.} titanium (Ti$\sim$ 20\,nm) and gold (Au$\sim$ 100\,nm) were thermally deposited in a thermal evaporator at a background pressure less than 1$\times$10$^{-7}$\,mbar.

The device was subjected to rapid thermal annealing at a temperature of 300\textdegree C for 1\,min. The unwanted areas, which create electrical contact between the micro-heater and the active area, were then selectively etched through oxygen plasma ashing. Positive photoresist S1813 was used as etch mask for this dry etching process. Figure\,\ref{Fig.1}(a) shows the scanning electron microscopic image of a part of the device in the vicinity of the micro-heater. Panel (b) of figure\,\ref{Fig.1} shows the schematic depiction of the device. Probe 1 and 4 ($1^{\prime}$ and $4^{\prime}$) were used as current probes while 2 and 3 ($2^{\prime}$ and $3^{\prime}$) are the voltage probe for the four-probe resistivity measurement for the two metal lines, which serve as resistive-thermometers at the two locations. Contact pads 1 and $1^{\prime}$ were also used to perform 2-probe current (\textit{I}) vs voltage (\textit{V}) measurements. \textit{I-V}  profiles were recorded using Keithley-6487 picometer-voltage source. Two lock-in amplifiers, Signal Recovery-7225 and Stanford Research-SR830 (phase-locked with each other) were used to measure the 4-probe resistances (say, $R_{th}$ and $R^{\prime}_{th}$) of the thermometers. A storage-type liquid nitrogen cryostat was used to perform electric and thermoelectric measurements at different temperatures ranging from 80 to 420\,K. The chamber was also utilized for in situ annealing of the device in vacuum.  Resistance vs temperature calibration of both the micro-thermometers was carried out by slowly increasing the cryostat temperature from 80\,K to 420\,K. Ramp rate of approximately 1\,Kmin$^{-1}$ was used during calibration to avoid  temperature lag between the sensor located in the cold finger (Pt-100) and the device. After the calibration one of the lock-in amplifier (Signal Recovery-7225) was reconfigured to measure the temperature difference between the micro-thermometers of the device. Special care was taken to minimize the common-mode gain across the on-chip thermometers. Keithley-6221 current source was used to excite the micro-heater. Thermoelectric voltage across the device was picked up using Keithley-2182a nanovoltmeter, across terminal 1 and 1$^{\prime}$.

\begin{figure}[t]
	\centering
	\includegraphics[scale=1]{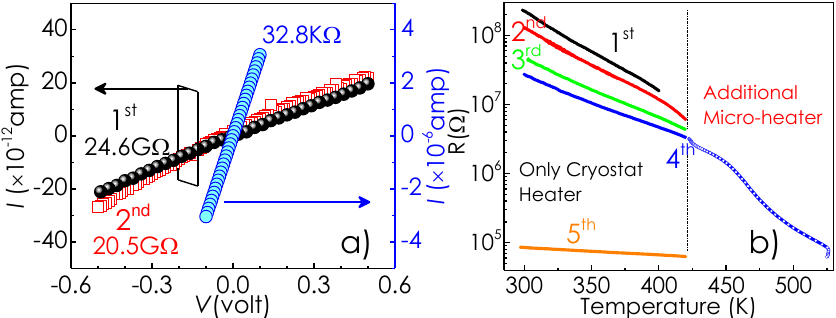}
	\caption{(a) Current-voltage  characteristics recorded at room temperature for the sample in different resistive states. (b) Resistance (\textit{R})  versus temperature plots recorded after different stages of annealing. Number written beside respective profiles represents the order of the annealing cycle. Temperature variation up to the dotted mark was achieved by an external heater while the on chip micro-heater was additionally used to reach higher temperature values.}
	\label{Fig.2}
\end{figure}

Room temperature \textit{I} versus \textit{V} profiles for the device are shown in figure\,\ref{Fig.2}(a). Black symbols represent the data recorded before evacuating the sample space of the cryostat. Resistance is measured to be $\sim$26\,G$\Omega$. The profile represented by red symbols is obtained after evacuating the sample space to $\sim 1 \times 10^{-5}$\,mbar. Clearly, evacuation at room temperature has no significant effect on the resistance of the device. Also, note that $I$-$V$ profiles are quite linear at such a highly resistive state. Next, the sample is heated using a cartridge heater embedded in the sample holder of the cryostat. Panel (b) shows a set of resistance ($R$) versus temperature ($T$) data recorded during successive annealing of the sample.
Evidently, at every step,  $R$ decreases with increasing $T$, which is expected for semiconductors. However, interestingly  $R$ does not go back to the same value when the temperature is reduced to room temperature after every step of annealing. Rather it follows a lower path to reach a smaller value. For example, room temperature resistance of the device before initiating any annealing process is recorded to be $\sim$230\,M$\Omega$. After the 1$^{\text {st}}$ and 2$^{\text {nd}}$ runs, $R$ decreases to 128\,M$\Omega$ and 44\,M$\Omega$, respectively. At the forth annealing step,  once the sample temperature is reached 420\,K, the highest value achievable using the embedded heater, the micro-heater fabricated on the chip is switched on. Current through the micro-heater is increased to 75\,mA in steps of 1\,mA . These data are marked as 4$^{\text {th}}$. It has been found that the micro-heater can rise the device temperature to 525\,K. Interestingly, the resistance of the device shows several orders of magnitude reduction after this high temperature annealing. $R$ vs $T$ profile  recorded at  5$^{\text{th}}$ step is also shown in figure\,\ref{Fig.2}(b). Clearly, at this step the rate of reduction of $R$ with increasing $T$ is much less than that is found in all the previous annealing steps.  Room temperature $I$ -$V$ profile for this lowest resistive state is also plotted in figure\,\ref{Fig.2}(a). At this state,  resistance is measured to be 32.8\,k$\Omega$, which is about six orders of magnitude less than the resistance measured before the sample goes through any annealing treatment. It has been further noticed that the resistance is practically unchanged even after weeks in vacuum. These findings provide an opportunity to arrest the resistive state of the device at any desired value. 

\begin{figure}[h]
	\centering
	\includegraphics[scale=1]{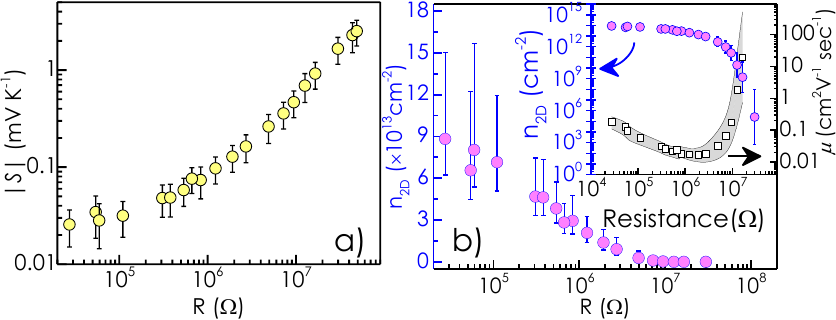}
	\caption{(a) Magnitude of Seebeck coefficient ($|S|$) as a function of device resistance ($R$). (b) Electron concentration ($n_{2D}$) obtained from $S$ as a function of $R$. Inset shows the variation of  $n_{2D}$ and mobility ($\mu$) in log-log plots.}
	\label{Fig.3}
\end{figure}

Change in resistance after annealing can occur either due to the variation in carrier concentration or mobility or both. In order to track the change of these parameters, we perform thermoelectric measurements at room temperature after taking the device to different resistive states. We start with Seebeck coefficient ($S$) measurement of the sample when it is at the lowest resistance state. Later, the sample space is successively purged with air to take the resistance to a  higher value. Once a desired resistive state is achieved, the chamber is evacuated to $\sim1 \times 10^{-5}$\,mbar before carrying out the thermoelectric measurement. $S$ is found to be negative in all cases, which suggests that electrons are the majority carriers in this system (n-type). Magnitude of $S$ as a function of the resistance is plotted in figure\,\ref{Fig.3}(a). Clearly,  $|S|$ increases with $R$. 
Current under a temperature gradient in a semiconductor can be expressed as $I$ = $G \Delta V$ + $G_S \Delta T$\cite{TERASAKI}, where $G$ the conductance of the sample, $\Delta T$ and $\Delta V$ are the temperature and potential differences between the contacts. In open circuit configuration, since $I$ = 0, above equation leads to $\Delta V$ = -$(G_s/G) \Delta T$ = $S \Delta T$. Seebeck coefficient $S$ can thus be defined as the ratio between the diffusion current ($G_S \Delta T$) and the product of the conductance $G$ and $\Delta T$. While the diffusion current depends upon the difference of carrier densities ($\delta n$) at the hot and the cold ends of the device, the conductance $G$ is proportional to the average carrier concentration $n$. Therefore, it can be said that the Seebeck coefficient is somewhat proportional to the ratio of these two quantities ($\delta n$ and $n$). In an open circuit configuration, the diffused charges accumulate on the colder side, which results in an upward shift of the Fermi level with respect to that of the warmer end. However, such a redistribution of carriers also leads to a potential difference between the two ends (cooler-end turns higher in potential for the electrons as compared to the hotter-end). A steady-state is finally achieved when this built-in potential restricts any further diffusion to take place. This situation is schematically depicted in figure\,\ref{Fig.1}(c). The ratio $\delta n$/$n$ is expected to decrease as the background carrier concentration $n$ in the layer increases. As a result, $|S|$ decreases with resistance. An analytical expression for $S$ can be obtained by solving linearized Boltzmann equation for $G_s \text{ and } G$\cite{npj_QM_TE}, which for 2D-semiconductors takes the form\cite{prb_thermoelectric}:
\begin{equation}
S=-\frac{k_B}{q_e}\left[\eta-\frac{(2+r)\int_0^{\infty} f^o \, \epsilon ^{r+1} \, d\epsilon}{(1+r)\int_0^{\infty} f^o \epsilon ^r \, d\epsilon}\right]
\label{eq:Seebeck Coeffi}
\end{equation}
where, $k_B$ the Boltzmann constant and $q_e$ the electron charge, $f^o$ the Fermi distribution function $[f^o=1/{1+exp(\epsilon-\eta)}]$. $r$ is known as the scattering exponent that depends upon the type of scattering mechanism limiting the carrier relaxation time. In \mos, acoustic phonons are found to be the most dominant scatters for the electrons at room temperature. Note that for acoustic phonon scattering, $r$ = 0\cite{prb_thermoelectric} in 2D. Equation\,\ref{eq:Seebeck Coeffi} is solved to obtain $\eta=(E_f-E_c)/k_BT$ and subsequently, 2D electron concentration $n_{2D}$ is estimated from $n_{2D}=8\pi m^* k_B T h^{-2} ln (1+e^{\eta})$, where $m^*$ and $h$ represent the electronic effective mass and Planck's constant, respectively. 
Figure\,\ref{Fig.3}(b) shows the variation of $n_{2D}$ as function of device resistance $R$. As expected the carrier concentration of the sample decreases with increasing resistance. A lower bound estimate of mobility, $\mu$ can be made from $n_{2D}$ and $R$  through $\mu=l/(bRn_{2D}q_e)$, where $l$ and $b$ are the length and width of the device. Inset of figure\,\ref{Fig.3}(b) portrays both $\mu$ and $n_{2D}$ as functions of $R$ in logarithmic scale. It is interesting to note that beyond a threshold value of $R$, both the parameters show a sharp change. $n_{2D}$ decreases by several orders of magnitude and at the same time $\mu$ increases significantly.    

\begin{figure}
	\centering
	\includegraphics[scale=1]{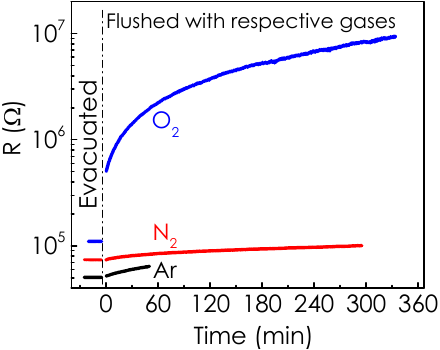}
	\caption{Change in Resistance of the device after exposing the sample in different environments. Conductivity of the channel reduces drastically after introduction of O$_2$ to the sample space.}
	\label{Fig.4}
\end{figure} 
All these observations point towards the fact that environment surrounding these 1L-\mos films plays an important role in governing their resistances. For better understanding, we have measured resistance of the device under different controlled environments. Results are shown in figure\,\ref{Fig.4}. The sample is first annealed and taken to a state where the resistance is approximately 50\,K$\Omega$. Sample space is then flushed with argon (Ar) gas (99.999\% pure, oxygen $\le$ 2\,PPM, moisture $\le$ 2\,PPM). Clearly, there is no significant increment in the device resistance as a result of the Ar-exposure. About an hour later, the sample space is evacuated. The same procedure is repeated with nitrogen (N$_2$) (99.999\% pure, oxygen $\le$ 2\,PPM, moisture $\le$ 2\,PPM). In this case, as well the change in resistance is not significant, which is obvious from the fact that even after $\sim$5\,hr of exposure, the resistance increases only by a factor of 1.3. However, the resistance changes dramatically when oxygen (99.999\% pure, nitrogen $\le$ 5\,PPM, moisture $\le$ 2\,PPM)  is finally introduced into the chamber. Within a few minutes, $R$ is increased by an order of magnitude and after 5.5\,hr, $R$ enhances by about two orders of magnitude.

\begin{figure*}
	\centering
	\includegraphics[scale=1]{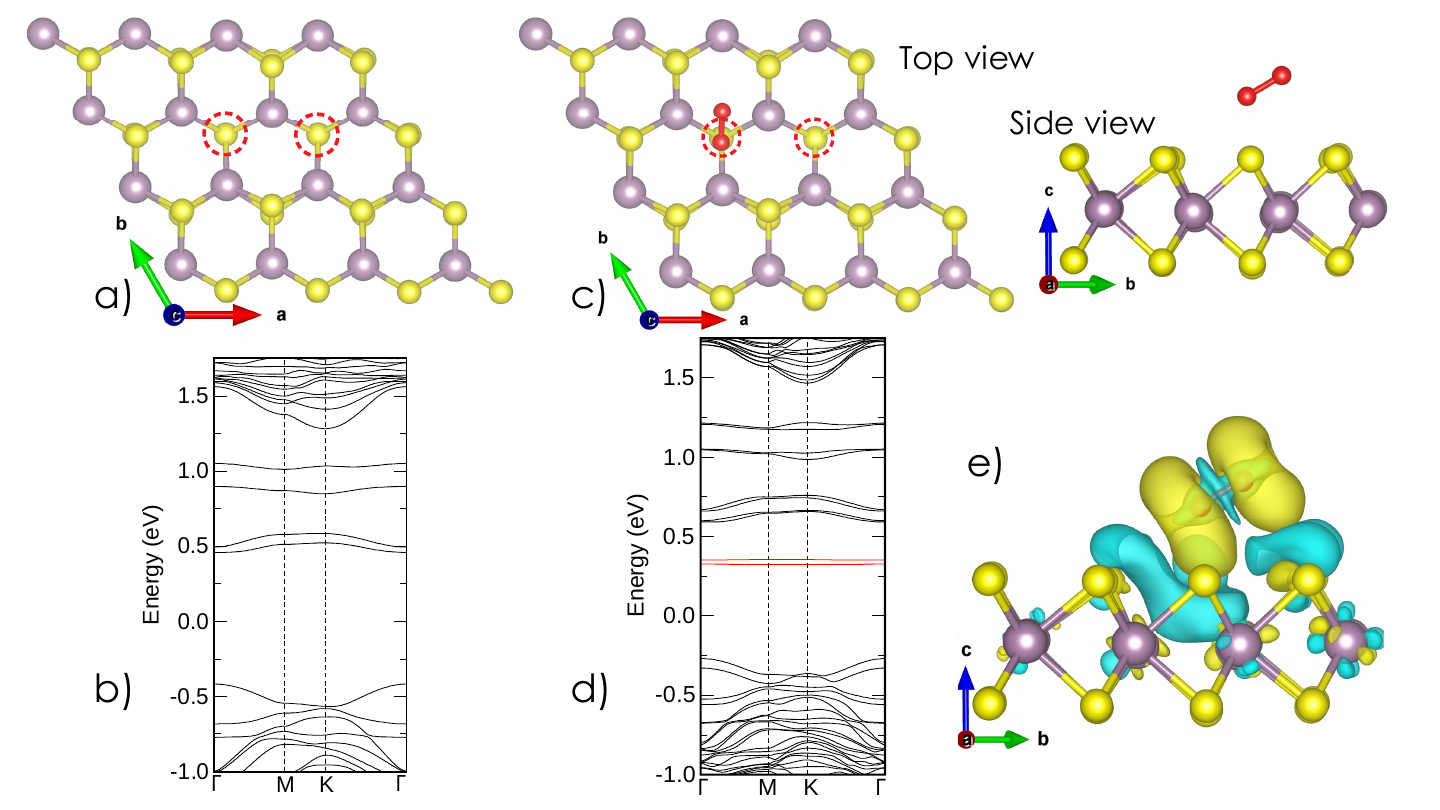}
	\caption{(a) The relaxed geometry (4$\times$4 supercell) and (b) band structure of the \mos monolayer with SV$_2$. (c) The relaxed geometry and (d) band structure of the complex system (O$_2$ adsorbed \mos monolayer with SV$2$). The red-colored bands in the gap region are predominantly contributed by the O$_2$ molecule. (e) Side view of the  differential charge density plot of the complex system with an isovalue of 9.8$\times$10$^{-5}$ e\AA$^{-3}$, where the yellow color indicates the electron-rich region and the cyan color denotes the loss of electrons.}
	\label{Fig.5}
\end{figure*} 
Observations of figure\,\ref{Fig.3}\,and\,\ref{Fig.4} clearly suggest that adsorption/desorption of oxygen molecules at the surface must be the reason for the dramatic change in carrier density of the layer. It should be noted that sulfur vacancies are shown to be the most abundant type of point defects, especially in CVD grown \mos. Among different types of sulfur vacancies in a monolayer film, same side di-sulfur vacancies, $SV_2$ (absence of two sulfur ions from the same side of a unit cell) are expected to act as donors\cite{defectdop1,defectdop2}. We believe that at the ambient condition, adsorption of oxygen molecules at these vacancy sites passivates the donors resulting in the reduction of the electron concentration in the conduction band. Upon vacuum annealing, these molecules are removed from the surface and the electron concentration recovers. The range of temperature involved in these experiments as well as the reversible nature of the phenomenon suggests that the oxygen molecules must be physisorbed (not chemisorbed) on the surface. In order to check the feasibility of this hypothesis theoretical calculations have been carried out within the framework of density functional theory\cite{Kohn_et_al} by employing the Vienna Ab-initio Simulation Package (VASP)\cite{Kresse_et_al,Kresse_2_et_al} software. For the purpose, we used projector augmented wave (PAW) pseudo-potentials\cite{Blochl_et_al,Kresse_et_al} with a kinetic energy cut-off of 450 eV, and Perdew-Burke Ernzerhof (PBE) exchange-correlation functional\cite{Perdew_et_al}. To optimize the structures, we employed a k-mesh of 5$\times$5$\times$1, while for the density of state (DOS) calculations, 21$\times$21$\times$1 k-grid were used. Total energy and force convergence criteria were chosen to be 10\textsuperscript{-6}\,eV and 0.02\,eV\AA{}$^{-1}$, respectively. The effects of spin-polarization and spin-orbit coupling were also included in our calculations. In this work, the van der Waals force correction was incorporated using DFT-D3 method\cite{Grimme_D3}. For the relaxed structure, the pressure in the supercell was less than 0.2 Kbar. At least 20\,\AA{} vacuum was introduced along the \textit{c}-direction to minimize the spurious interactions. To perform the band structure calculations, 60 k-points were considered in the reciprocal space. A 4$\times$4 supercells were considered to understand the nature of the interaction of O$_2$ molecules with the sulfur-vacancies in \mos monolayer. Two sulfur atoms are removed from same side of the unit cell as shown in figure\,\ref{Fig.5}(a). Calculations show that the vacancy $SV_2$ acts as a donor state and give rise to several new energy states in the forbidden gap as presented in panel (b) of figure\,\ref{Fig.5}. Note that the top of the donor band lies approximately 0.25\,eV below the conduction band minimum (CBM). These donor states appear due to the hybridization of weak sulfur 3p and strong molybdenum 4d orbitals. It should be noted that the finding is in good agreement with the theoretical predictions made earlier\cite{Qiu_et_al}. In order to look into the effect of O$_2$ adsorption at the surface, we first carried out geometry relaxation with the O$_2$ molecule placed at an initial guess position. The energetically most favorable site for attachment comes out to be on the top of any vacancy site of the defect and $\sim$0.19\,nm above from the monolayer surface. The top and the side views of the complex system (1L-\mos with $SV_2$+O$_2$) are depicted in figure\,\ref{Fig.5}(c). In figure\,\ref{Fig.5}(d), band structure calculated for $SV_2$+O$_2$ complex is shown. Attachment of O$_2$ creates two flat bands (red lines) that are approximately 250~meV below the donor levels.  Rest of the band structure remains to be almost the same as that is obtained for 1L-\mos with $SV_2$. The charge density difference in O$_2$ adsorption can be described as, $\rho_{ad}=\rho_{MoS_{2}+SV_2+O_{2}}-\rho_{MoS_{2}+SV_2}-\rho_{O_{2}}$, where $\rho_{MoS_{2}+SV_2+O_{2}}$, $\rho_{MoS_{2}+SV_2}$, and $\rho_{O_{2}}$ are the charge densities of the 1L-\mos with \textit{SV$_2$}+adsorbed O$_2$, 1L-\mos with \textit{SV$_2$}, and isolated O$_2$ molecule, respectively. The differential charge density plot of the complex system is presented in figure\,\ref{Fig.5}(e). Note that upon adsorption of O$_2$ molecule, the 1L-\mos loses electrons, while these are accumulated around O$_2$ molecule suggesting electron transfer from $SV_2$ defects to O$_2$ molecule, which is also consistent with the appearance of energy levels (red lines) below the $SV_2$ donor states when O$_2$ is adsorbed at the $SV_2$ defect site as shown in figure\,\ref{Fig.5}(d). Theory thus supports the picture that the adsorption of O$_2$ molecules passivates $SV_2$  donor states by introducing a lower energy state that traps these electrons. The formation energy of $SV_2$+O$_2$ complex is estimated to be $\sim$340\,meV suggesting a weak bonding of the O$_2$ molecules with the surface. This the reason why at elevated temperatures and in high vacuum, O$_2$ molecules are efficiently detached from the surface that results in a  drastic increase in its conductivity as observed experimentally.



Adsorption of oxygen molecules at the sulfur vacancy sites results in the formation of electron traps, which lie  250\,meV below the donor levels introduced by these vacancies in 1L-\mos layers.  As a result, these donors are passivated and the as-grown layer that is exposed to the ambient condition becomes highly resistive.  It has been found that the resistance, as well as the thermoelectric power of the material, can be tuned to a large extent by controlling the temperature of annealing and oxygen exposure. Once established, these parameters remain unchanged unless the layer is exposed to oxygen environment. Annealing followed by an effective encapsulation method can offer a unique way to control doping in the material, which is an important milestone towards technology application. Moreover, these findings generate excitements to use 1L-\mos for oxygen gas sensing applications. 


We acknowledge the financial support by Council of Scientific \& Industrial Research of Government of India under the project Code:03(1403)/17/EMR II. We thank Centre of Excellence in Nanoelectronics -CEN of IIT Bombay, for providing various experimental facilities.
\bibliography{Man_ref}

\end{document}